%% file: main.tex
\documentclass[a4paper]{IEEEtran}
\usepackage[noadjust]{cite}
\usepackage{graphicx} 
\usepackage{caption,subcaption}
\usepackage[english]{babel}
\usepackage{multirow}
\usepackage{bm}
\usepackage{mathtools}
\usepackage[square, numbers, sort]{natbib}
\usepackage{kantlipsum} 
\usepackage{mwe} 
\usepackage{amsmath, amsthm, amssymb}
\newtheorem{theorem}{Theorem}

\newtheorem{corollary}{Corollary}
\usepackage[font=small,labelfont=bf]{caption}
\usepackage[dvipsnames]{xcolor}
\usepackage[colorlinks]{hyperref}
\hypersetup{
linkcolor=BrickRed
,citecolor=Green
,filecolor=Mulberry
,urlcolor=NavyBlue
,menucolor=BrickRed
,runcolor=Mulberry
,linkbordercolor=BrickRed
,citebordercolor=Green
,filebordercolor=Mulberry
,urlbordercolor=NavyBlue
,menubordercolor=BrickRed
,runbordercolor=Mulberry
}

\DeclareRobustCommand*{\IEEEauthorrefmark}[1]{%
  \raisebox{0pt}[0pt][0pt]{\textsuperscript{\footnotesize #1}}%
}
\begin{document}

\title{Over-the-Air Federated Learning with Enhanced Privacy}

\author{
\IEEEauthorblockN{
    Xiaochan Xue\IEEEauthorrefmark{a}\IEEEauthorrefmark{1}, 
    Moh Khalid Hasan\IEEEauthorrefmark{a}\IEEEauthorrefmark{1}, 
    Shucheng Yu\IEEEauthorrefmark{a}\IEEEauthorrefmark{1}, 
    Laxima Niure Kandel\IEEEauthorrefmark{b}\IEEEauthorrefmark{2}, 
    Min Song\IEEEauthorrefmark{a}\IEEEauthorrefmark{1}}\\
\IEEEauthorrefmark{a}\IEEEauthorblockA{Department of Electrical and Computer Engineering, Stevens Institute of Technology, NJ 07030}\\
\IEEEauthorrefmark{b}\IEEEauthorblockA{Electrical Engineering and Computer Science Dept, Embry-Riddle Aeronautical University, FL 32114}\\

Email:\IEEEauthorrefmark{1}\{xxue2, mhasan12, syu19, msong6\}@stevens.edu, \IEEEauthorrefmark{2}\{Laxima.NiureKandel\}@erau.edu
}
\maketitle

\begin{abstract}
Federated learning (FL) has emerged as a promising learning paradigm in which only local model parameters (gradients) are shared. Private user data never leaves the local devices thus preserving data privacy. However, recent research has shown that even when local data is never shared by a user, exchanging model parameters without protection can also leak private information. Moreover, in wireless systems, the frequent transmission of model parameters can cause tremendous bandwidth consumption and network congestion when the model is large. To address this problem, we propose a new FL framework with efficient over-the-air parameter aggregation and strong privacy protection of both user data and models. We achieve this by introducing pairwise cancellable random artificial noises (PCR-ANs) on end devices. As compared to existing over-the-air computation (AirComp) based FL schemes, our design provides stronger privacy protection. We analytically show the secrecy capacity and the convergence rate of the proposed wireless FL aggregation algorithm.
\end{abstract}


\input{1_intro}

\input{2_system_model}

\input{3_design}

\input{4_analysis}

\input{5_evaluation}

\input{6_conclusion}


\footnotesize
\bibliographystyle{unsrtnat}
\bibliography{main}

\end{document}

%% file: 1_intro.tex
\section{Introduction}

In machine learning, especially deep learning, large-scale collection of sensitive data entails both high bandwidth consumption and privacy-related risks. To mitigate these limitations and leverage the power of proliferating edge devices, federated learning (FL) \citep{mcmahan2017communication} has emerged as a promising new learning paradigm. 
In FL each edge device trains a local ML model using its private data and uploads only model parameters to a central server. The server then aggregates local models received from the distributed edge devices to obtain a global model that is expected to outperform the individual local models. %
While FL is promising as compared to centralized learning, the frequent transmission of model parameters can still cause significant bandwidth consumption and latency in wireless and mobile systems. Moreover, recent research has discovered vulnerabilities of FL under membership inference attacks \citep{hayes2017logan, shokri2017membership, melis2019exploiting, wei2020framework}. Specifically, it has been demonstrated that models implicitly memorize inappropriate details about the underlying training data and can reveal sensitive information to attackers inadvertently.
To strike a balance between efficiency and privacy in FL, existing research has resorted to various techniques, including  
\emph{Secure Aggregation} (SA) \citep{bonawitz2017practical} and \emph{Differential Privacy} (DP) \citep{dwork2008differential}. 
The former obfuscates parameters to the aggregator but needs pairwise key exchange which incurs non-trivial communication costs in edge computing environments. The latter on the other hand injects random noises into local training data so that it is computationally indistinguishable from that of other individuals. For FL, local differential privacy (LDP), which is a mode of DP, is more suitable because of its distributed nature and users can add noises to model parameters locally before disclosing them to the \textit{untrusted model aggregator}. 
While LDP has the advantage of lower computational and communication overheads, it poses its own challenges. Specifically, an LDP model needs to introduce noises at a significantly higher level than what is required in a DP model. Also, since each user perturbs its parameters individually, the aggregated variance highly depends on the number of users participating in the training \citep{bassily2017practical}. 

Recently, the feasibility of over-the-air computation (AirComp) \citep{4305404} coupled with LDP is being explored within the context of FL, to overcome communication bottlenecks and provide additional protection to local model privacy. The AirComp-based approach exploits the broadcast and the natural superposition property of wireless multiple access channels (MAC) for fast, free, and more efficient global model aggregation. The key idea is the simultaneous synchronized transmission of linear-analog modulated local gradients. With appropriate pre-channel coefficient equalization, superposed RF signals over the air can be demodulated as the additive result at the receiver without actually performing the addition operation. Together with local pre-processing, complex functions such as scalar products can be implemented via AirComp, which saves both local computation and latency for wireless devices.   
Despite of the challenges, existing research \citep{8849334, 8970161, hellstrom2020wireless, 8870236} has demonstrated promising progresses both theoretically and through practical implementation. 
Along this direction, this paper aims to explore the full potential of AirComp-based FL by providing enhanced privacy protection. Specifically, while protecting model privacy, existing research \citep{8849334, 8970161, hellstrom2020wireless, 8870236, 9174426} mainly relies on \emph{obfuscation via aggregation} (OVA) of parameters from multiple users with local adjustment of signal to noise ratio (SNR). Although this approach protects model privacy against the parameter aggregation server (PAS), such protection is fragile under stronger attack models in which an external attacker is equipped with a directional antenna to overhear RF signals from individual transmitters and bypass the aggregation. Moreover, the OVA approach requires a higher noise level for model privacy when the number of users is less, which adversely impacts the global model quality. To address this limitation, in this paper we introduce pairwise cancellable random artificial noises (PCR-ANs) to obfuscate individual private model parameters. By adjusting the PCR-AN level, our design is able to thwart external eavesdroppers equipped with directional antennas. Because the PCR-ANs are pairwise cancellable, only residue noises remain in the aggregated model. Our design can be considered as a novel integration of SA and DP at the physical layer. Analytical results provide both secrecy capacity and the FL convergence rate of our design. Our contributions can be summarized as follows:

\begin{itemize}

\item We introduce a new AirComp-based privacy-preserving FL scheme considering the presence of powerful eavesdroppers. The pairwise cancellable random artificial noise (PCR-AN) design leverages the properties of both secure aggregation and differential privacy and provides a better trade-off between privacy and model utility as compared to the state-of-the-art.
\item We theoretically analyze the feasibility of the PCR-AN design and formulate the secrecy capacity of our proposed privacy-preserving FL scheme in the presence of powerful eavesdroppers. We analytically show the convergence rate of our proposed FL scheme.
\item With the adjustable power parameters of artificial noises, our design is also able to preserve model privacy at the PAS based on the differential privacy constraints.
\end{itemize}

The rest of the paper is structured as follows. Section \ref{System Model} describes the system and threat model for FL. Section \ref{design} presents our design and elaborates on PCR-AN-aided privacy-preserving FL. Section \ref{analysis} presents an analytical privacy analysis, the secrecy capacity, and the convergence rate of our proposed scheme. Section \ref{Evaluation} presents the simulation, and evaluation results, and Section \ref{conclusion} concludes the paper.

%% file: 2_system_model.tex
\section{System Model and Assumptions} \label{System Model}

\subsection{System Model and Federated Learning}
We consider a wireless federated learning system consisting of a parameter aggregation server (PAS) and multiple end users. The PAS is a single-antenna receiver and aggregates the distributed local model parameters from total $K$ ($K=|\mathcal{K}|$) users, where $\mathcal{K} = \{1,2,3,...,2i\},i\in\mathbb{Z}^+$. 
Each user participant $k$ ($k \in \mathcal{K}$) is a spatially distributed single-antenna device and without loss of generality, it is assumed that all devices are identical to each other and within one-hop distance to PAS. Each user $k$ has a private local data set $\mathcal{D}_k$ and we assume that all users have the same data size of $|\mathcal{D}_k|$. Data points are denoted as 
$\mathcal{D}_k=\{(\boldsymbol{u}^{(k)}_j, v^{(k)}_j)|j \in \mathcal{D}_k\}$, where
$\boldsymbol{u}^{(k)}_j \in \mathbb{R}^d$ is the $j$-th data point and $v^{(k)}_j$ is the corresponding label for each data point.
Each user individually trains an ML model using their private data $\mathcal{D}_k$ and then uploads a $d$-dimensional model parameter vector $\boldsymbol{\mathrm{w}}$ wirelessly to the PAS. 
For efficiency, the participants use Gaussian multiple access channels (MAC) to simultaneously transmit their respective parameters. PAS receives aggregated parameters because of the over-the-air superposition of wireless signals. This process is called the over-the-air computation (AirComp) \citep{4305404} which can implement complex functions if users are well synchronized and equalized. 
The global aggregated model is obtained by  minimizing the loss function $F(\boldsymbol{\mathrm{w}})$ as follows:
\begin{equation}  \label{eq: loss function}
    \boldsymbol{\mathrm{w}}^* = \arg \min_{\boldsymbol{\mathrm{w}}}F(\boldsymbol{\mathrm{w}})\triangleq \frac{1}{|\mathcal{D}|}\displaystyle\sum_{k=1}^{K}\displaystyle\sum_{j=1}^{\mathcal{D}_k}f_k((\boldsymbol{u}^{(k)}_j, v^{(k)}_j);\boldsymbol{\mathrm{w}})
\end{equation}
where $\mathcal{D}=\displaystyle\bigcup_{k=1}^{K}\mathcal{D}_k$ denotes the entire dataset used for training, and $f_k(\bullet)$ is the loss function for user $k$. The minimization of $F(\boldsymbol{\mathrm{w}})$ in eq. (\ref{eq: loss function}) is carried out iteratively through a gradient descent (GD) algorithm. At iteration $t$, the PAS broadcasts the global model parameter vector $\boldsymbol{\mathrm{w}}_{t}$ and each user then updates its local gradient vector over the local dataset $\mathcal{D}_k$ as:

\begin{equation}
    \boldsymbol{g}_{k}( \boldsymbol{\mathrm{w}}_{t})= \frac{1}{|\mathcal{D}_k|}\displaystyle\sum_{j=1}^{\mathcal{D}_k}\nabla f_k((\boldsymbol{u}^{(k)}_j, v^{(k)}_j);\boldsymbol{\mathrm{w}})
\end{equation}
Next, the locally computed gradient is sent back to the PAS and the global model $\boldsymbol{\mathrm{w}}_t$ is updated according to: 

\begin{equation}\label{eq: global model}
    \boldsymbol{\mathrm{w}}_{t+1} = \boldsymbol{\mathrm{w}}_{t}-\eta_t \frac{1}{K} \left(\displaystyle\sum_{k=1}^{K} \boldsymbol{g}_{k}( \boldsymbol{\mathrm{w}}_{t})\right)
\end{equation}
$\boldsymbol{\mathrm{w}}_{t+1}$ is the updated global model and $\eta_t$ is the learning rate of the GD algorithm at iteration $t$. The PAS will broadcast $\boldsymbol{\mathrm{w}}_{t+1}$ and the above process continues until convergence with total $T$ iterations.

\subsection{Threat Model}
Our threat model considers honest-but-curious attackers, i.e., we assume the attacker passively eavesdrops on exchanged messages (e.g., gradients) between the client and the PAS. However, the attacker does not interfere with the training process. For instance, due to the broadcast nature of the wireless medium, the eavesdropper easily wiretaps the local parameter-modulated transmitted signal by pointing a directional antenna toward the transmitting victim device. After the adversary has wiretapped model at its disposal, it can violate privacy by recovering the underlying sensitive data on which the model was trained by launching sophisticated model inversion attacks or may gain leaked private information when the wiretapped model is used for inference. We show that our design defends against such passive attackers and achieves both data and model privacy. More sophisticated active attackers will be explored in our future work.



%% file: 3_design.tex
\section{Our Design} \label{design}
\subsection{Preliminaries of AirComp for Ultrafast Aggregation}
AirComp shows great promise to support ultrafast aggregation of local FL model parameters from  distributed mobile users. 
The principle idea of AirComp is to exploit the analog-wave superposition property of wireless multiple access channels (MAC). 
As illustrated in Fig. \ref{fig: AirComp}, we consider a simplified baseline single-antenna AirComp system with non-zero receiver noise and unequal channel coefficients. Let $s_k$ denote the analog modulated local model parameters symbols calculated by client $k$. The aggregated function at the PAS then can be written as:
\begin{figure}[ht]
    \centering
    \includegraphics[width=90mm]{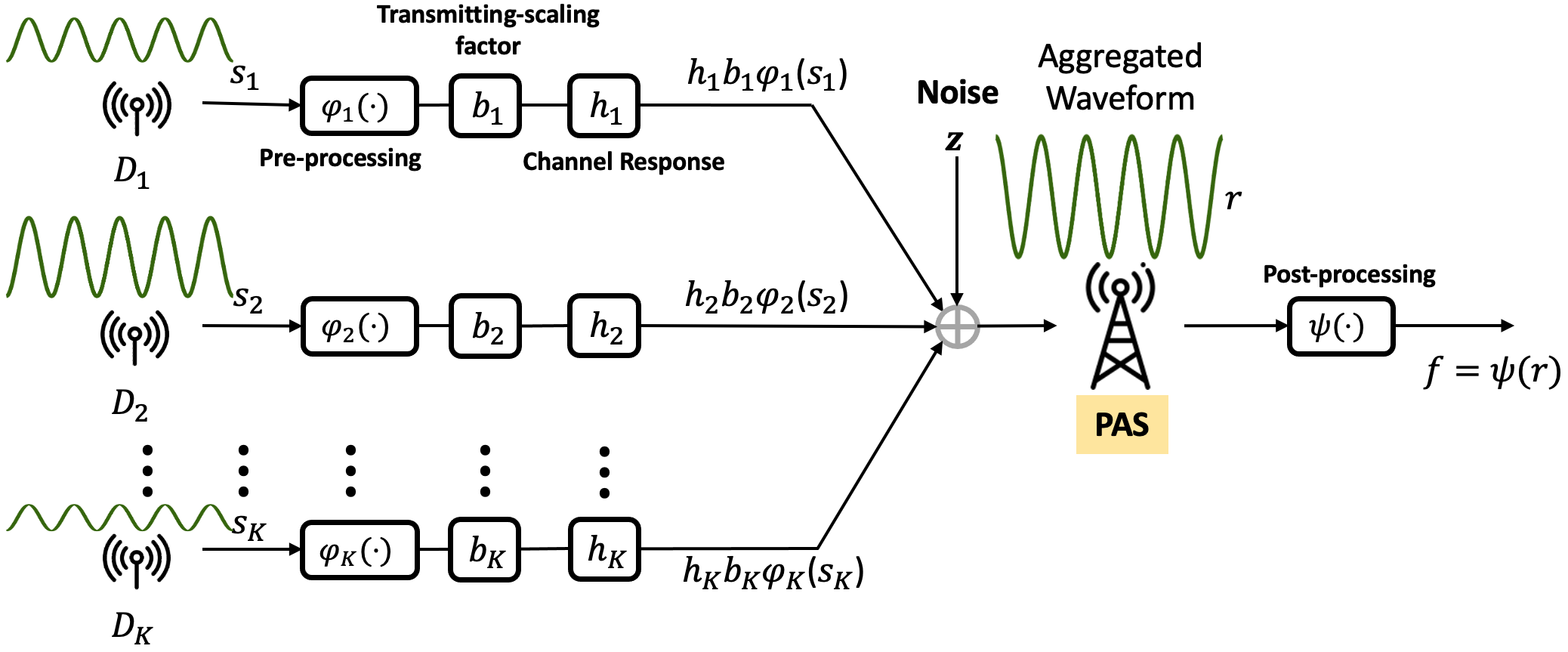}
    \caption{The data aggregation over the MAC via over-the-air computation.}
    \label{fig: AirComp}
\end{figure}

\begin{equation}
f=\psi(r)
 \label{eq: received sum}
\end{equation} 
\begin{equation}
 r= {\sum_{{k=1}}^{K}h_kb_k\varphi_k(s_{k})}+z_k
\end{equation}
where in eq. (\ref{eq: received sum}), $r$ is the superimposed received signal and $\psi(\bullet)$ is the post-processing function at the PAS. $\varphi_k(\bullet)$ is the pre-processing function at each transmitting device. The selection of pre-processing and post-processing functions depends on the desired function $f$. The variable $h_k$ is the channel coefficient, $b_k$ is the transmitter scaling factor to achieve channel inversion (CI) and $z_k$ is the Additive White Gaussian Noise (AWGN) at user $k$. It is assumed that the channel is time-invariant and the transmitting mobile devices including the PAS have the channel state information (CSI) to achieve channel inversion.

\subsection{Pairwise Cancellable Random Artificial Noise (PCR-AN)}
\begin{figure}[t]
    \centering
    \includegraphics[width=80mm]{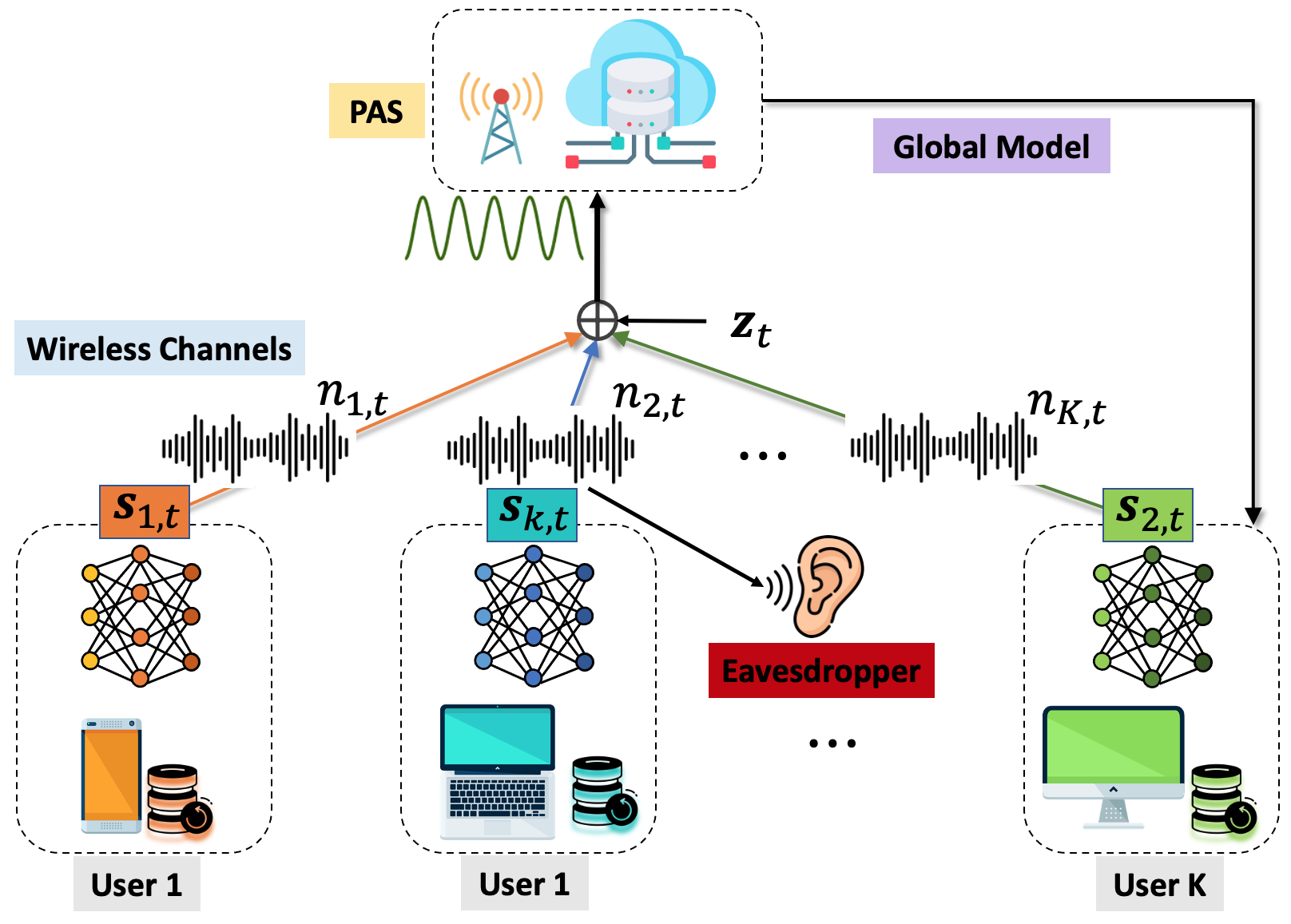}
    \caption{Federated learning with artificial noises based on AirComp in the presence of an eavesdropper.}
    \label{fig: FL + AirComp}
\end{figure}

We present a general gradient aggregation scheme for wireless FL based on AirComp, as shown in Fig. \ref{fig: FL + AirComp}. Each user $k$ synchronously transmits a linear combination of local gradients and pairwise cancellable random artificial noise (PCR-AN) over a wireless channel for total $T$ training iterations. At each iteration $t$, all participating $K$ users transmit their local computed gradient vector $\boldsymbol{s}_{k,t}:=\boldsymbol{g}_k(\boldsymbol{\mathrm{w}}_t) \in \mathbb{R}^{d}$ masked with 
PCR-AN to preserve modal privacy. More specifically, the transmitted signal of user $k$ with added artificial noise $\boldsymbol{n}_{k,t}$ at iteration $t$ is given as:
\begin{equation} \label{eq: transmit signal}
\boldsymbol{x}_{k,t}=b_k \varphi_k \left( \boldsymbol{s}_{k,t} + \boldsymbol{n}_{k,t} \right)+ \boldsymbol{z}_{k,t}
\end{equation}
The terms used in eq. (\ref{eq: transmit signal}) are explained below:
\begin{itemize}
    \item $\boldsymbol{n}_{k,t} \in \mathbb{R}^{d}$ is the PCR-AN (Gaussian noise) with mean $\mu_{k,t}$ , and variance $\sigma_{k,t}^2$ $(\boldsymbol{n}_{k,t} \sim \mathcal{N}(\mu_{k,t},\sigma_{k,t}^2))$ to mask the gradient vector, $\boldsymbol{s}_{k,t}$. Two pairwise devices secretly share the mean and variance value, then add artificial noise with opposite mean values to the gradients. For example, users $a$ and $b$ pre-share a secret ($\mu,\sigma^2$) and then this secret will be used by user $a$ to add noise of  $ \mathcal{N}(+\mu,\sigma^2_a)$ and noise of  $ \mathcal{N}(-\mu,\sigma^2_b)$ is added by user $b$.
    \item $\boldsymbol{z}_{k,t} \in \mathbb{R}^{d}$ is the additive zero-mean unit-variance Gaussian noise over the wireless channel ( $\mathcal{N}(0,\sigma_z^2)$, $\sigma_z^2=1$).
    \item $\varphi_k(\bullet)$ is the pre-processing at each user. Since, the desired function at the PAS in the context of FL is the arithmetic mean, $\varphi_k(\bullet)=1$. 
    \item $b_k$ is the Tx-scaling factor for each user to ensure the analog modulated waves add constructively in the air and a non-zero signal is received. Typically, the signal is multiplied by $e^{-j\phi_k}$ for local phase correction.
\end{itemize}

Also, in the above eq. (\ref{eq: transmit signal}), it is assumed that the gradient vectors have a bounded norm to bound the maximum changing rate, i.e., $\|\boldsymbol{s}_{k,t}\|_2 \leq L_s, \forall k$. Let $\alpha_k \in [0,1]$ denote the coefficient of power dedicated to the gradient vector $\boldsymbol{s}_{k,t}$. The remaining power of $\beta_k \in[0, 1-\alpha_k]$ ($\beta_k \geq \alpha_k$) is dedicated to the artificial noise to satisfy the maximum transmit power constraint needs of $P_k$. Using eq. (\ref{eq: global model}) to (\ref{eq: transmit signal}), the received signal at the PAS can be written as:
\begin{equation} \label{eq: received signal}
    \boldsymbol{r}_t = \displaystyle\sum_{k=1}^{K} |h_k|\left( \frac{\sqrt{\alpha_k P_k}}{L_s} \boldsymbol{s}_{k,t} + \sqrt{\beta_k  P_k}\boldsymbol{n}_{k,t} \right)+\boldsymbol{z}_{k,t}
\end{equation}

To represent eq. (\ref{eq: received signal}) in compact form, we introduce $m$ as follows:
\begin{equation} \label{eq: m}
    m:=|h_k|\frac{\sqrt{\alpha_k P_k}}{L_s},\forall k
\end{equation}
Herein, $m$ is a constant, and the upper bound of $m$ can be computed by utilizing $\alpha_k \leq 1, \forall k$ in eq. (\ref{eq: m}). To maximize the power of aligned gradients, $m$ is chosen as $m = \frac{\sqrt{\min\limits_{q}|h_q|^2P_q}}{L_s}$ resulting in $\alpha_k$ as follows:
\vspace{-1.5em}
\begin{center}  
\[ \alpha_k = \frac{\min\limits_{q}|h_q|^2P_q}{|h_k|^2P_k} \] 
\end{center}
$q$ is the user with worst effective SNR. Thus, above choice of $\alpha_k$ shows that the alignment of gradients is effectively limited by the user $q$ with the worst effective SNR. Substituting $m$ in eq. (\ref{eq: received signal}), we get the compact representation as follows: 
\begin{equation} \label{eq: received signal_simple}
    \boldsymbol{r}_t = m\displaystyle\sum_{k=1}^{K} \boldsymbol{s}_{k,t} + \displaystyle\sum_{k=1}^{K} |h_k|\sqrt{\beta_k  P_k}\boldsymbol{n}_{k,t} +
    \boldsymbol{z}_t
\end{equation}

As seen in eq. (\ref{eq: received sum}), the PAS performs post-processing on received signal $\boldsymbol{r}_t$ and for the aggregation scheme, the post-processing function is $\psi(\bullet)=\frac{1}{mK}$. Thus, the estimated function at PAS is as follows:
\begin{equation} \label{eq: estimated}
\begin{split}
    \hat{\boldsymbol{s}_t} & = \frac{1}{m K} (\boldsymbol{r}_t) \\
    & =\underbrace{\frac{1}{K} \displaystyle\sum_{k=1}^{K} \boldsymbol{s}_{k,t}}_\text{$\nabla F(\mathrm{w}_t)$} + 
    \underbrace{\frac{1}{m K} \displaystyle\sum_{k=1}^{K} |h_k|\sqrt{\beta_k  P_k} \boldsymbol{n}_{k,t} }_\text{$A_t$} +
    \frac{1}{m K} \boldsymbol{z}_t
\end{split}
\end{equation}
where $A_t + \frac{1}{m K} \boldsymbol{z}_t$ is the effective noise at the PAS. Since the pairwise devices add artificial noise of opposite mean, the summed artificial noise and channel noise will have a mean of $0$ and variance of $ \frac{1}{m K} \displaystyle\sum_{k=1}^{K} |h_k|\sqrt{\beta_k  P_k} \boldsymbol{n}_{k,t}+ \frac{1}{m K} \boldsymbol{z}_t$. Therefore, the PAS receives an unbiased estimate of the average gradient  $\nabla F(\boldsymbol{\mathrm{w}}_t)$.

%% file: 4_analysis.tex
\section{Analysis} \label{analysis}
In this section, we first evaluate the privacy protection provided when local wireless devices participating in the same learning task obfuscate local parameters through PCR-ANs. We show that the additive artificial noise protects individual users' privacy without interfering with the global model aggregation at PAS. Next, we discuss the secrecy capacity of our design in the presence of an eavesdropper who is listening to the user's communication with the PAS. Lastly, we prove the proposed FL scheme is convergent and show the optimization of convergence, which can also meet the differential privacy requirement to preserve privacy at PAS. 

\subsection{PCR-AN Aided Privacy} \label{Artificial Noise for Privacy}
As mentioned in Section \ref{design}, we allocate higher power to PCR-ANs such that SNR is low and the sensitive data is below the noise floor. This means any malicious device eavesdropping over the air can only acquire noise instead of sensitive data. However, in prior literature, low SNR would mean difficulty reconstructing the original data at PAS. Herein, we expatiate the feasibility of our design despite low SNR; we present a detailed analysis showing the added PCR-AN will not interfere with the reconstruction at PAS.

Let $i$ represent the $i$-th pair of wireless devices $(+i,-i)$, where $+i \in \frac{+\mathcal{K}}{2}$ denotes the device adding a positive mean value of artificial noise, and $-i \in \frac{-\mathcal{K}}{2}$ denotes the device adding a negative mean value of artificial noise. Note, $\left(\frac{+\mathcal{K}}{2}\right)\cup \left(\frac{-\mathcal{K}}{2}\right)=\mathcal{K}$ and $\left(\frac{+\mathcal{K}}{2}\right) \cap \left(\frac{-\mathcal{K}}{2}\right)=0$. 
The mean values of added artificial noises at user $+i$ and user $-i$ are also pairwise, i.e., user $+i$ and $-i$ adds $\boldsymbol{n}_{+i,t}= \mathcal{N}(\mu_{+i,t},\,\sigma^{2}_{+i,t})$ and $\boldsymbol{n}_{-i,t}= \mathcal{N}(\mu_{-i,t},\,\sigma^{2}_{-i,t})$, respectively. The PCR-ANs are randomly selected by users to mask the uploading gradient vector. Thus, summed PCR-ANs, denoted as $A_t$ in eq. (\ref{eq: estimated}) can be written as:
\begin{equation}
    A_t := \frac{1}{m K}(\displaystyle\sum_{i=1}^{K/2}|h_i|\sqrt{\beta_i  P_i} \boldsymbol{n}_{i,t} +\displaystyle\sum_{i=-1}^{-K/2}|h_i|\sqrt{\beta_i  P_i} \boldsymbol{n}_{i,t})
\end{equation}
\begin{equation} \label{eq: At}
    A_t := \frac{1}{m K}(\displaystyle\sum_{i=1}^{K/2}|h_i|\sqrt{\beta_i  P_i}) (\underbrace{\displaystyle\sum_{i=1}^{K/2} \boldsymbol{n}_{i,t} +\displaystyle\sum_{i=-1}^{-K/2} \boldsymbol{n}_{i,t}}_\text{Cancellable Noise (CN)})
\end{equation}
\begin{equation} \label{eq: CN1}
    CN:=  \displaystyle\sum_{i=1}^{K/2} \mathcal{N}(\mu_{+i,t},\,\sigma^{2}_{+i,t})+\mathcal{N}(\mu_{-i,t},\,\sigma^{2}_{-i,t})
\end{equation}
\begin{equation} \label{eq: CN2}
    CN:= \displaystyle\sum_{i=1}^{K/2} \mathcal{N}\left(0,\,(\sigma^{2}_{+i,t}+\sigma^{2}_{-i,t})\right)
\end{equation}
Above eq. (\ref{eq: CN1}) to eq. (\ref{eq: CN2}) is based on the the property of PCR-ANs i.e., $(\mu_{+i,t}+\mu_{-i,t})=0$. Therefore, aggregated PCR-ANs at PAS will follow the distribution $\mathcal{N}(0,\sigma_A^2)$, where $\sigma_A^2 = \displaystyle\sum_{k=1}^{K/2} (\sigma^{2}_{+k,t}+\sigma^{2}_{-k,t})$. The aggregated variance $\sigma_A^2$ is bounded by the \emph{Central Limit Theorem} (CLT). The uploading gradient for each user includes numerous parameters, which indicates the convergence in aggregated variance $\sigma_A^2$ from \textbf{Corollary \ref{corollary: aggregated variance}}.
\begin{corollary} \label{corollary: aggregated variance}
    All added artificial noises are independent but not identically distributed. The $\mu_k$ and $\sigma_k^2$ for each user $k$ satisfy the \textbf{Lyapunov’s Condition}. Therefore, according to \textbf{Lyapunov’s Central Limit Theorem}, the distribution of aggregated variances $\sigma_A^2$ of all artificial noises is convergent.
\end{corollary}
The proof of Lyapunov’s central limit theorem is out of the scope of this paper and interested readers in the proof and Lyapunov's condition are referred to \citep{alma991004198139705251,10.1145/3481646.3481652}. The high-power PCR-AN added to each user with different distributions will not interfere with the PAS to reconstruct the aggregation of locally trained model signals. The estimated function from eq. (\ref{eq: estimated}) can be written as:
\begin{equation} \label{eq: estimated_without Artificial noise}
     \hat{\boldsymbol{s}_t} = \underbrace{\frac{1}{K}\displaystyle\sum_{k=1}^{K}\boldsymbol{s}_{k,t}}_\text{$\nabla F(\mathrm{w}_T)$}  +
     \underbrace{\frac{1}{m K} \left(\displaystyle\sum_{k=1}^{K} {|h_k|} {{\sqrt{{\beta_{k}} {P_{k}}}} {\boldsymbol{n}_{k,t}}} + {\boldsymbol{z}_{t}}\right)}_{\text{$\boldsymbol{z}_{t}^{'}$}}
\end{equation}

We denote $\frac{1}{m K}(\sum_{i=1}^{K/2}|h_i|\sqrt{\beta_i  P_i})$ in eq. (\ref{eq: At}) as $M$, and $\boldsymbol{z}_t^{'}\sim\mathcal{N}(0,\sigma_{\boldsymbol{z}_t^{'}}^2)$ is the residual noise of aggregated PCR-ANs and channel noise at PAS, where $\sigma_{\boldsymbol{z}_t^{'}}^2=M^2 \cdot{\sigma_{A}}^{2}+{\sigma_{z}}^{2}$. As $\boldsymbol{z}_t^{'}$ is zero mean, $\hat{\boldsymbol{s}_{t}}$ is an unbiased estimate of $\nabla F(\boldsymbol{\mathrm{w}}_T)$.

\subsection{Secrecy Capacity} \label{Secrecy Capacity}

To estimate the secrecy capacity, we select a two-user scenario. Herein, we consider two pairwise users $a$ and $b$. User $a$ transmits its signal with added PCR-AN of mean of $\mu_{a,t}$, and user $b$ transmits the parameters with added PCR-AN of mean of $\mu_{b,t}$. The signals of both users add in the air and the server receives the sum of the user signals. The signal-to-noise ratio of the received sum signal ($SNR_s$) at the PAS is given by
\begin{equation}
    SNR_s = \frac{{\frac{{\sqrt {{\alpha _a}{P_a}} }}{{{L_s}}}{{\left| {{h_a}} \right|}^2}}}{\sigma_{\boldsymbol{z}_t^{'}}^2}  
\end{equation}
\noindent where ${\sigma _A^2}$ is the residual noise of PCR-ANs after the aggregation at the server. The capacity at the server for user $a$ can be represented as
\begin{equation}
\begin{split}
    {C_s} 
    & = {\log _2}\left( {1 + SN{R_s}} \right) \\
    & = {\log _2}\left( {\frac{{\sqrt {{\alpha _a}{P_a}} }}{{{L_s}}}{{\left| {{h_a}} \right|}^2} + \sigma_{\boldsymbol{z}_t^{'}}^2} \right) - {\log _2}\left( \sigma_{\boldsymbol{z}_t^{'}}^2 \right)
\end{split}
\end{equation}
\indent We assume the eavesdropper wiretaps the data of user $a$. As the eavesdropper receives the PCR-AN with the actual signal from user $a$ with variance $\sigma_a^2$, the SNR at the eavesdropper is given by,
\begin{equation}
{SN{R_{ev}}} = \frac{{\frac{{\sqrt {{\alpha _a}{P_a}} }}{{{L_s}}}{{\left| {{{h_a^{\left( e \right)}}}} \right|}^2}}}{{\sigma _z^2 + \sigma _a^2}}
\end{equation}
\noindent where ${{{\left| {h_a^{\left( e \right)}} \right|}^2}}$ is the channel power gain corresponding to the channel coefficient ${h_a^{\left( e \right)}}$. The capacity at the eavesdropper is estimated as follows:

\begin{align}
\label{capacity_eq}
    {C_{ev}} = {\log _2}\left( {\frac{{\sqrt {{\alpha _a}{P_a}} }}{{{L_s}}}{{\left| {h_a^{(e)}} \right|}^2} + \sigma _z^2 + \sigma _a^2} \right) - {\log _2}\left( {\sigma _z^2 + \sigma _a^2} \right)
\end{align}

Now, we can estimate the secrecy capacity as follows.

\begin{equation}
\label{sec_cap}
    C = {\left[ {{{\log }_2}\left( {\frac{{\frac{{\sqrt {{\alpha _a}{P_a}} }}{{{L_s}}}{{\left| {{h_a}} \right|}^2} + \sigma_{\boldsymbol{z}_t^{'}}^2}}{{\frac{{\sqrt {{\alpha _a}{P_a}} }}{{{L_s}}}{{\left| {h_a^{\left( e \right)}} \right|}^2} + \sigma _z^2 + \sigma _a^2}}} \right) - {{\log }_2}\left( {\frac{\sigma_{\boldsymbol{z}_t^{'}}^2}{{\sigma _z^2 + \sigma _a^2}}} \right)} \right]^ + }
\end{equation}

\noindent where ${\left[ x \right]^ + } = \max \left\{ {x,0} \right\}$. The main objective is to enhance the secrecy capacity so that the privacy of user $a$ is improved. It can be realized from eq. (\ref{sec_cap}) that there is a direct influence of ${\sigma _a^2}$ on $C$, which means that more PCR-AN at the sender decreases the capacity at the eavesdropper, in other words, increases the privacy of user $a$. Also, it is evident in eq. (\ref{sec_cap}) that if ${\sigma _A^2}$ increases, $C$ decreases.

\subsection{Convergence Rate of Private AirComp-based FL}
\begin{theorem} \label{theorem: convergence}
Suppose the loss function $F$ is $\lambda$-strongly convex and $\mu$-smooth with respect to $\boldsymbol{\mathrm{w}}^*$ over a convex set $\mathcal{W}$, and $\mathbb{E}[\|\hat{\boldsymbol{s}_t}\|^2] \leq G^2$. Then if we pick $\eta_t=1/\lambda_t$, the convergence rate for iteration $T$ is
\begin{equation}\label{eq: SGD_specific}
\begin{split}
    & \mathbb{E}\left[ F(\boldsymbol{\mathrm{w}}_T)-F(\boldsymbol{\mathrm{w}}^*)\right]\\
    & \leq \frac{2\mu}{\lambda^2 T} \left( L_s^2 + \frac{d}{m^2 K^2}\left[\sum_{k=1}^{K}|h_k|^2\beta_k P_k + \sigma_z^2\right]\right)    
\end{split}
\end{equation}
\end{theorem}
The detailed proof of Theorem 1 is given by \citep{rakhlin2011making} and \citep{9174426}. The convergence rate can also be maximized by optimizing the artificial noise parameter $\beta_k$, which can also meet the differential privacy requirement to preserve the privacy at PAS in \citep{9174426}. The $\beta_k$ can be written as:
\begin{equation}
    \beta_k = \frac{Z_k}{{|h_k|}^2P_k}, \forall k
\end{equation}
where $Z_k=\min\left[ \lambda_k, (\Psi-\displaystyle\sum_{p=1}^{k-1}U_p)^+ \right], \forall k$, $\Psi = \max\limits_{p} \frac{\min\limits_{q}|h_q|^2P_q}{\epsilon_p}\log\frac{1.25}{\delta}-\sigma^2_z$, and $U_p = {|h_p|}^2 \beta_p P_p$. The $(\epsilon, \delta)$ is the local differential privacy level.




%% file: 5_evaluation.tex
\section{Evaluation} \label{Evaluation}
In this section, we first provide simulation results of secrecy capacity to show the performance of our AirComp-based privacy-preserving FL model. The Rayleigh fading wireless channels for the simulation results are randomly generated over $10^6$ realization samples in Matlab. The channel coefficients are drawn from $\mathcal{N}(0,1)$, and the channel noise variance is set to $\sigma_z^2=1$. We set the variance $\sigma_k^2$ of the user $k$'s PCR-AN to 25dB. The Lipschitz constant $L_s$ is considered as 1. 
\begin{figure}[t]
    \centering
    \includegraphics[width=3.3in]{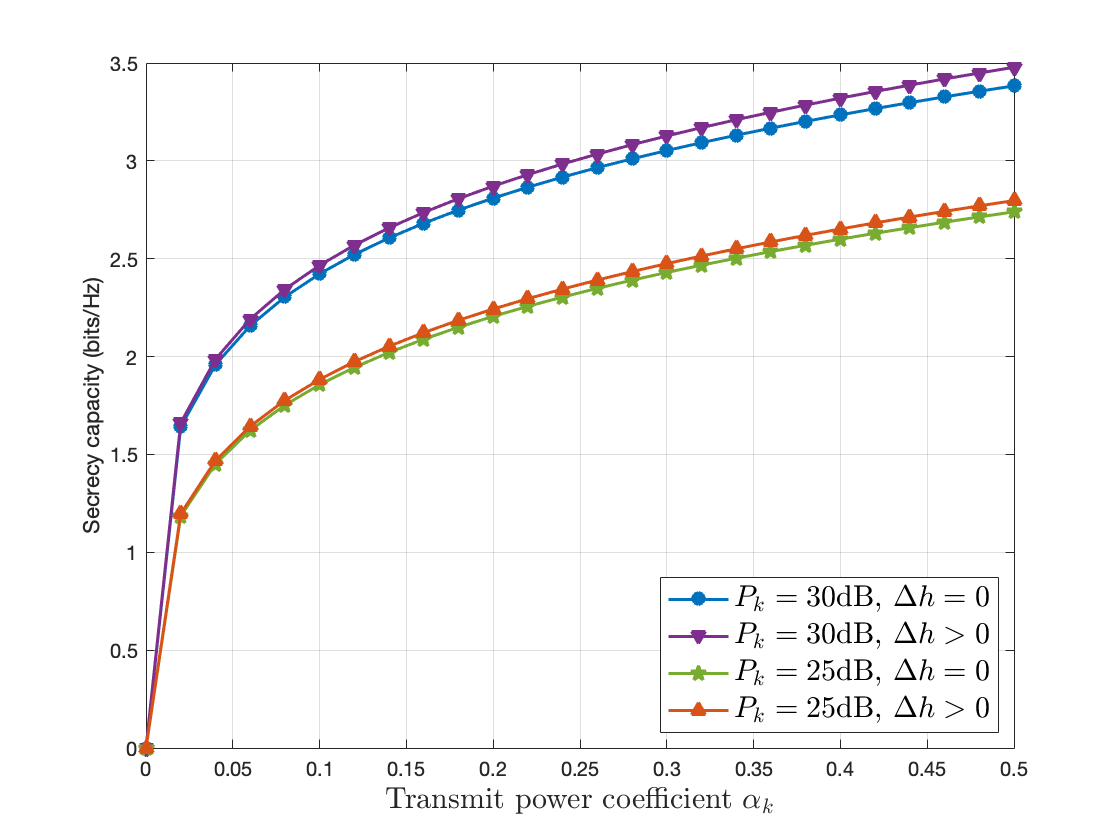}
    \caption{Secrecy capacity with respect to different transmitted signal's coefficient $\alpha_k$.}
    \label{fig: simulation-Secrecy capacity alpha}
\end{figure}
\begin{figure}[b]
    \centering
    \includegraphics[width=3.3in]{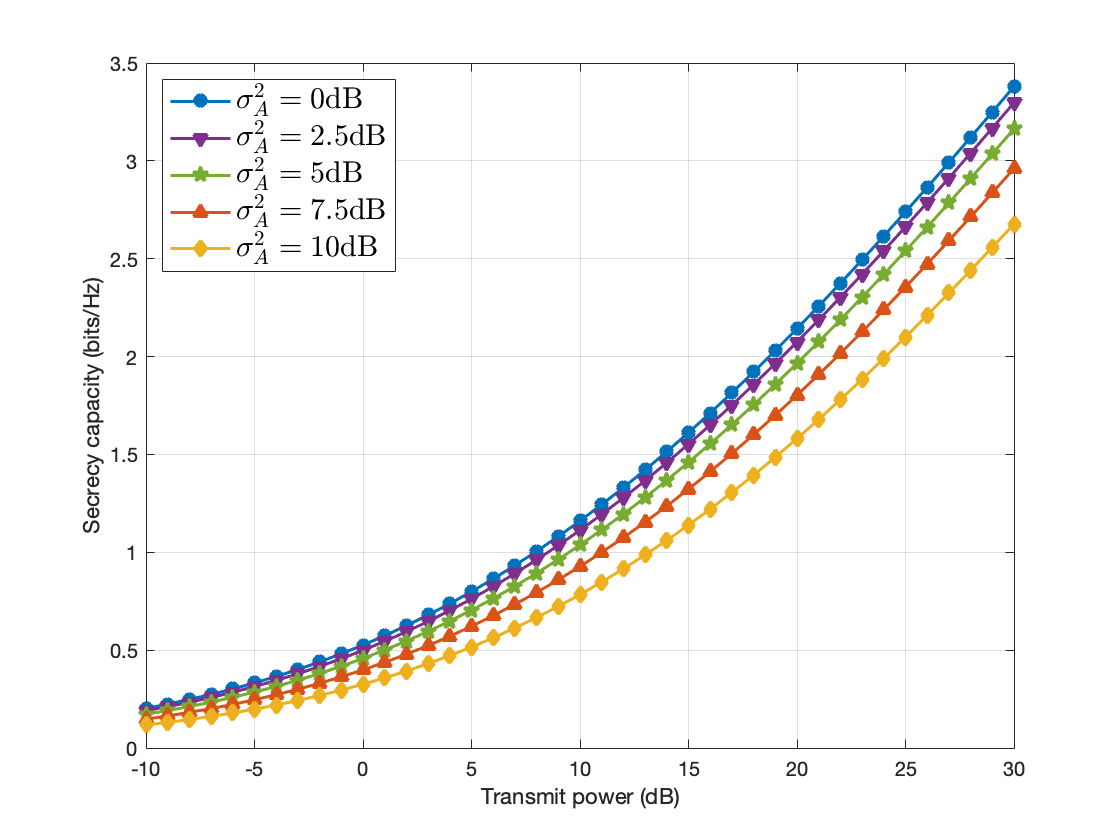}
    \caption{Secrecy capacity with respect to different transmit power.}
    \label{fig: simulation-Secrecy capacity}
\end{figure}
We define $\Delta h = |h_k|^2 - |h_k^{\left(e \right)}|^2$ as the capability of an eavesdropper to obtain the parameters from the victim compared with PAS. $\Delta h = 0$ means the eavesdropper has high capability as PAS, $\Delta h > 0$ means low capacity at eavesdropper, relatively. Based on the assumption of two coefficients $\alpha_k$ and $\beta_k$, we show the secrecy capacity with the respect to different transmit signal's coefficient $\alpha_k \in[0,0.5]$ in Fig. \ref{fig: simulation-Secrecy capacity alpha}. With the increase of transmit power coefficient $\alpha_k$, the secrecy capacity increases for all scenarios. For both transmit power $P_k=25$ dB and $P_k=30$ dB, the secrecy capacity increases for the scenario that the eavesdropper has a worse channel gain ($\Delta h > 0$) than PAS.

We then consider the $\Delta h > 0$ scenario here, which is a general assumption in wireless communication. We set the $\alpha_k = 0.5$ in the simulation. In Fig. \ref{fig: simulation-Secrecy capacity}, we show the impact of transmit signal power on the secrecy capacity of an individual user. For aggregated PCR-ANs variance $\sigma^2_A = 0$ dB, which means the power of PCR-ANs is cancelled perfectly at PAS. With the convergence of aggregated PCR-ANs' variance $\sigma_A^2$, the secrecy capacity increase. The trend of secrecy capacity also increases with the increase of transmit power.

\begin{figure}[t]
    \centering
    \includegraphics[width=3in]{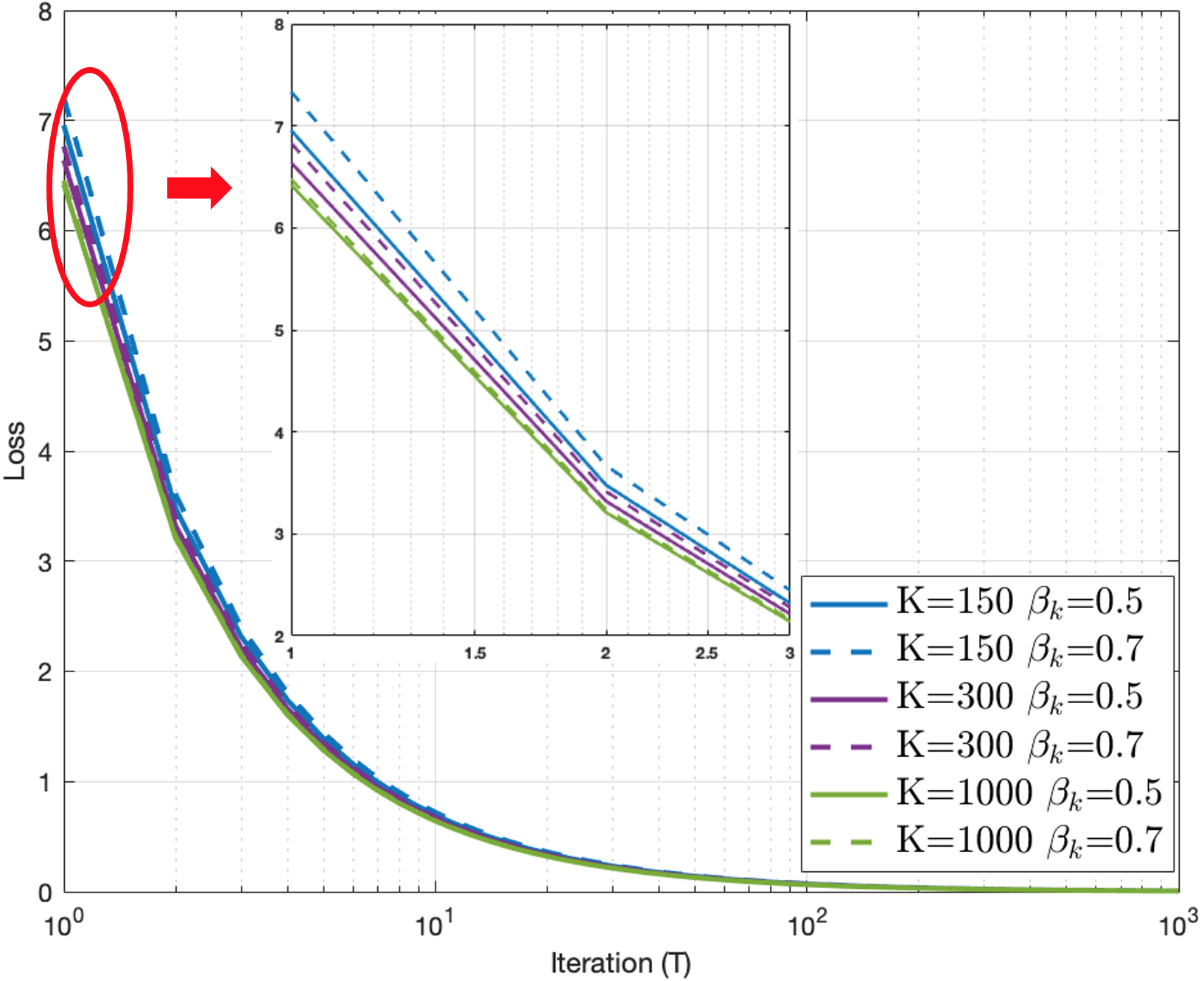}
    \caption{Convergence rate of private AirComp-based FL.}
    \label{fig: convergence rate}
\end{figure}
Fig. \ref{fig: convergence rate} shows the impact of the total number of users $K$ and iteration $T$ on the convergence rate based on eq. (\ref{eq: SGD_specific}). For the GD algorithm, the regularization parameter $\lambda$ is $10^{-3}$ and $T = 1000$ training iterations. We assume the transmit power $P_k=30$ dB for each user $k$ based on the analysis from Fig. \ref{fig: simulation-Secrecy capacity alpha} which can reach a higher secrecy capacity. We also assume the data points $d=30$. From the enlarged detail for the beginning of the iteration, as we increase the number of users, the training loss decays with $T$. We also show the impact of PCR-AN's power coefficient $\beta_k$ for each user $k$. The loss decreases with the decrease of PCR-AN's power. We compare different pair of coefficients of transmitting signal and PAC-AN, which is $\alpha_k=0.3$, $\beta_k=0.7$ and $\alpha_k=0.5$, $\beta_k=0.5$. From simulation results, the lower $\beta_k$ performs a faster convergence rate. Therefore, we chose to set $\beta_k=0.5$. This means only necessary PCR-AN power can help the FL model reach good convergence. We can easily figure out that the trend of training loss converges as the number of the iteration $T$ increases.

%% file: 6_conclusion.tex
\section{Conclusion}\label{conclusion}
In this paper, we propose a new privacy-preserving FL framework with efficient over-the-air parameter aggregation and random pairwise cancellable artificial noises (PCR-ANs) to obfuscate individual private model parameters. We demonstrate the use of PCR-ANs by users provides strong privacy protection for both user data and models.
By, adjusting the PCR-AN power level, our design is able to thwart external eavesdroppers equipped with directional antennas. Also, because the PCR-ANs are pairwise cancellable, it does not cause a large error in the estimation of the global model at the aggregator. Some residual noise due to different variances remains in the aggregated model which aids in providing additional protection against malicious servers. 
Theoretical analysis of the secrecy capacity and convergence rate shows the feasibility of our design and the stronger privacy protection provided by the proposed FL. 

\section*{Acknowledgment}
This work was supported in part by the National Science Foundation under grants ECCS-1923739 and CNS-1817438.